\DeclareUnicodeCharacter{2009}{\,}
\documentclass[a4paper,11pt]{article}
\usepackage{pos}
\usepackage{graphicx}
\usepackage{subcaption}
\usepackage{lineno}
\usepackage{siunitx}
\title{The {ATAR}, a dense and compact vertex detector for the {PIONEER} experiment}
\author*[a]{T. Rohe}
\onbehalf{for the {PIONEER} collaboration}

\affiliation[a]{Paul Scherrer Institut, \\ Forschungsstrasse 111, 5232 Villigen-PSI, Switzerland}
\emailAdd{tilman.rohe@psi.ch}

\abstract{
PIONEER is an approved experiment at the Paul Scherrer Institut (PSI) in Villigen, Switzerland that aims to measure the charged-pion branching ratio to electrons vs. muons with an accuracy of \SI{0.01}{\percent}. This is an order of magnitude improvement in precision compared to its predecessors and approaches the accuracy of the standard model prediction. The key elements of PIONEER are a high resolution calorimeter built concentrically around the high granularity active target (ATAR). The ATAR is composed of a stack of 48 low gain avalanche strip detectors (LGADs) to stop the beam pions and provide detailed information of the event topology. This will include precise timing (in the order of \SI{200}{ps}), spatial information (to an accuracy of about \SI{60}{\micro\metre}) and pulse height information with a high dynamic range to separate pion stops from muons and MIPs.
To avoid affecting the energy measurement of the positrons, the readout electronics is placed outside the acceptance angle of the calorimeter. It is composed of a fast amplifier and a fast sampling chip digitizing the LGAD wave form for real time analysis. The paper describes the PIONEER-experiment with focus on the ATAR and its readout. The status of the development and activities in 2026 will be reported.
}
\FullConference{
34th International Workshop on Vertex Detectors, Jul 20 – 24, 2026 in Stoss, Switzerland
}
\begin{document}

\maketitle

\section{Introduction}

The decay of the pion is the simplest electro-weak process involving quarks and can therefore be calculated with high precision. While the weak interaction is not sensitive to the lepton flavor, the decay of the pion to the positron ($\pi^{+} \rightarrow e^{+}\nu$) is helicity suppressed. The presently most accurate standard model predictions are presented in \cite{ref:theo1, ref:theo2, ref:theo3}:
    \begin{equation}
    R^{\pi}_{e/\mu} = \frac{\Gamma (\pi \rightarrow e\nu(\gamma))}{\Gamma (\pi \rightarrow \mu\nu(\gamma))} = (1.23524\pm  .00015) \times 10^{-4}
    \label{label:theoValue}
    \end{equation}
The accurancy of the calculation is in the order of \SI{0.01}{\percent} and exceeds the accuracy of the present measured world average by a factor of about 15:
    %\begin{equation}
    $R^{\pi}_{e/\mu} = (1.2327\pm .0023) \times 10^{-4}$ \cite{ref:pdg2024}.
    %\end{equation}
This average is dominated by the result of the PiENu-Collaboration published more than ten years ago $R^{\pi}_{e/\mu} = (1.2344\pm .0023 (stat) \pm .0019 (syst)) \times 10^{-4}$  \cite{PiENu:2015seu}.

Aim of the PIONEER collaboration \cite{ref:pioneer-prop} is to provide a measurement of the branching ratio with an accuracy comparable to the prediction to provide a probe of lepton universality. It is an approved experiment at the Paul Scherrer Institut (PSI) in Villigen, Switzerland and will be installed at the secondary beam line piE5.
In a later phase PIONEER will focus on the measurement of the pion beta decay $\pi^{+} \rightarrow \pi^{0}e^{+}\nu(\gamma)$ with a branching ratio of the order of $10^{-9}$ to provide a more accurate determination of the CKM matrix elements $\vert V_{ud}\vert$ and $\vert V_{us}\vert$.

\section{Measurement method}

The detector design is presently focused on the so called phase~I of the experiment where $R_{e/\mu}$ is determined. Both decay channels have a positron in the final state. However, they differ in their energy spectrum. The direct decay to a positron ($\pi^{+} \rightarrow e^{+}\nu$) has a narrow distribution around 69~MeV, as this is a two body decay, while the positrons from the dominating channel via the three body $\mu$-decay ($\pi^{+} \rightarrow \mu^{+}\nu(\gamma)$ and $\mu^{+} \rightarrow e^{+}\nu_{e}\bar\nu_{\mu}$) have a wide energy spectrum from 0 to 53~MeV. The branching ratio $R_{e/\mu}$ can in principle be determined by measuring the positron energy and counting the number of positrons above resp.\@ below an energy threshold of e.g.\@ 56~MeV.

%\marginpar{\bf\textcolor{red}{mention\\ rate}}
%However, the measured spectra broaden due to effects in the calorimeter (shower leakage, albedo, nuclear effects, etc.), interaction with material before entering the calorimeter, and pions resp.\@ mouns decaying in flight. Therefore the distributions overlap and elaborate analysis technics have to be developed to reach the targeted precision.

However, various detector effects cause events from one side to leak to the other side and therefore require elaborate analysis
techniques. Most notable, incomplete shower collection in the calorimeter (e.g. due to lateral leakage, albedo, nuclear effects
or interactions with material prior to reaching the calorimeter) can cause a \SI{69}{MeV} positron originating from a $\pi^{+} \rightarrow e^{+}\nu$
decay to be reconstructed below threshold energy. Simulations suggest that this affects slightly below \SI{1}{\percent} of all such
events. To reach the \SI{0.01}{\percent} precision goal, their exact number has to be determined at the \SI{1}{\percent} level.
This so called tail fraction was a limiting factor for the PiENu experiment \cite{PiENu:2015seu}.

%In order to obtain the required number of pion decays, a rate of $300\,$kHz pion stops inside the taget is necessary. This means that the average time between two pion stops is in the same magnitude as the live time of the muon. PIONEER needs to be constructed in a way that it can cope with this rate and the pile up of muon decays caused by this. Especially the low enery tail in the spectrum of the 69~MeV positrons, which can, according to simulations, be reduced to about 1\,\% of the events needs to measured with an accuracy of about 1\,\%.

Moreover, to reach the required number of events to meet the statistical uncertainty, running at a rate of \SI{300}{kHz} pion stops in the target
is anticipated. This means that the average time between two pion arrivals is on the same order of magnitude as the
muon lifetime, yielding a significant amount of cases where multiple pions decay chains are active simultaneously.
The segmented calorimeter and highly segmented ATAR are design choices made to cope with this rate requirement.

\section{The PIONEER apparatus}

%The two main components of the PIONEER-experiment are the high granularity fully active silicon target (ATAR) surrounded by a highly segmented calorimeter with good energy resolution.

The two main components of the PIONEER-experiment are the highly segmented calorimeter with good energy resolution and the fully instrumented target (ATAR) in the center.

\begin{figure}
    \centering
    \includegraphics[width=0.5\linewidth]{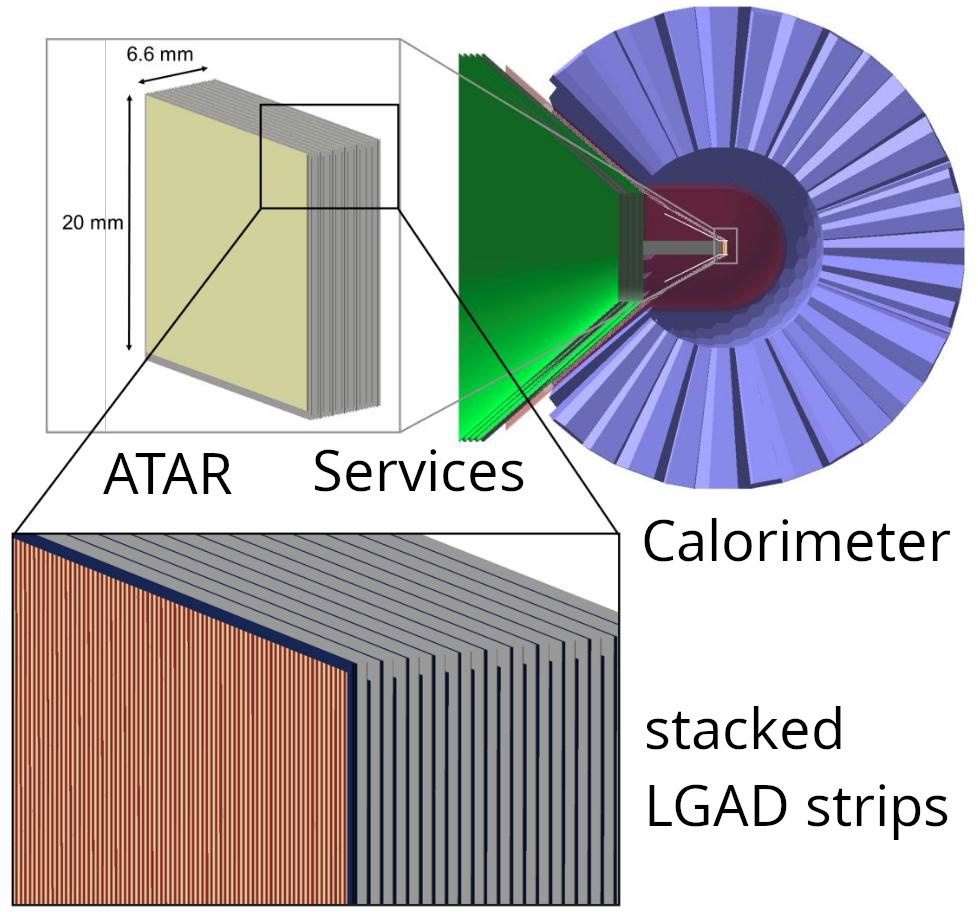}
    \caption{Schematic illustration of the Pioneer experiment and a close up of the target region}
    \label{fig:targetRegion}
\end{figure}

\subsection{Calorimeter}
The baseline for the calorimeter is to built it from Lutetium-Yttrium Oxyorthosilicate (LYSO) crystals, a calorimeter material with high light yield (32 000 photons/MeV), relatively fast decay time (40 ns), high density (7.4 g/cm$^{3}$), and high stopping power (radiation length $X_0 = $\SI{1.14}{cm} and Moli\`ere radius $R_M = $\SI{2.07}{cm}). A test beam experiment with six of those crystals at the PSI beamline piM1 in 2025 has has shown a very good performance. Especially the energy resolution (better than \SI{2}{\percent} for positron energies above \SI{40}{MeV}) and the time resolution (better than \SI{130}{ps} for positron energies above 30\,MeV) fulfill the requirements of the experiment \cite{ref:caloJINST}. An additional attractive feature of the crystal calorimeter is that it can be commissioned in stages. A first run of the experiment with a reduced angle coverage of the calorimeter is planned for the year after the shutdown for the PSI HIPA upgrade \cite{ref:HIMB}.

As a backup solution a liquid Xenon (LXe) calorimeter based on the technology developed for the MEG experiment \cite{ref:lXe} is also being considered.

\subsection{Active Target (ATAR)}
The most innovative feature of the PIONEER experiment is the fully instrumented target.
%Specify ATAR size 2cm x 2cm x 6.6 mm as well as strip sizes (2 cm x 200 um x 120 um) somewhere here.
It is $\SI{2}{cm}\times\SI{2}{cm}\times\SI{6.6}{mm}$ in size and consists on 48 stacked silicon LGAD strip sensors, read out by recording the full wave form of each of the 4800 strips.

LGADs have a very fast signal rise time which allows to measure the signal, e.g. the pion stop and its decay with a precision of 200\,ps.
By using very fast amplifiers and recording the full waveform, a pulse pair separation down to \SI{2}{ns} in the same channel can be reached e.g. a pion stop and the track of its decay product. This will provide the required capabilities to resolve
the tail fraction of $\pi^{+}\rightarrow e^{+}\nu (\gamma)$ events that fall below threshold. Considering
pion and muon lifetime (\(\tau_\pi \approx \SI{26}{ns}, \tau_\mu \approx \SI{2.2}{\micro\s}\)), selecting an
appropriate time window offers a rejection factor of 160 for the $\pi\rightarrow \mu\rightarrow e$ decay chain
while preserving around \SI{60}{\percent} of all \(\pi\rightarrow e\) events \cite{ref:sim}.
%The pion ($\tau_{\pi} \approx 26\,$ns) and muon ($\tau_{\mu} \approx 2200\,$ns) have life times differing by a factor of more than 80. Requiring a positron within an interval of 2-32\,ns after a pion stop, selects about 60\,\% of the direct decays ($\pi^{+} \rightarrow e^{+}\nu(\gamma)$) while at the same time rejecting positrons from a Michel decay ($\pi^{+} \rightarrow \mu^{+}\nu(\gamma)$ and $\mu^{+} \rightarrow e^{+}\nu_{e}\bar\nu_{\mu}$) by a factor of about 150 \cite{ref:sim}.

%The incoming 65\,MeV pion as well as the slow muon ionize about ten times more than a minimum ionizing particle (MIP) as the outgoing positrons. As the muon from the pion decay has only a low energy and a short range (total energy deposition about 4.1\,MeV), it can be distinguished from a pion (total energy deposition about 14\,MeV) from its total energy deposition. With suitable energy cuts a $\pi^{+} \rightarrow e^{+}\nu(\gamma)$ selection efficiency of about 80\,\% can be achieved while the decay channel via the muon is rejected by a factor of the order of 1000. This, however, requires a dynamic range of the pulse height measurement of more than 100.

Making use of the pulse height measurement, a stopping pion with a characteristic total energy deposit around \SI{14}{MeV} can be identified.
Similarly, the minimally ionising positron leaving ATAR has a very distinct energy distribution whereas the
stopping muon from a pion decay features a \SI{4.1}{MeV} energy deposit. Selecting events that only feature
a characteristic pion stop deposit followed by a matching positron deposit preserves \SI{80}{\percent} of the $\pi^{+}\rightarrow e^{+}\nu$
events while reducing the muon decay at rest background by 3 orders of magnitude in this tail reveal analysis.
%To efficiently detect the energy deposition of a MIP (about \SI{35}{keV} in \SI{120}{\micro\meter} of Silicon) and distinguish it from a pion and muon stop the analogue amplification chain needs a dynamic range above 100.

The full benefits of the topology reconstruction including per-hit energy deposits are still under investigation
but early results suggest that this offers another factor 500 rejection power while still retaining \SI{50}{\percent} of all $\pi^{+}\rightarrow e^{+}\nu$ events.
Moreover, the spatial reconstruction of the positron allows to identify the positron impact location in the
calorimeter and significantly contributes to pileup rejection and proper identification of radiative decays.
For the reconstruction of events a traditional rules based approach is followed as well as a more aggressive using transformer techniques \cite{ref:omar}.

%The spatial resolution of the strip detectors allow to reconstruct the event topology. The $\pi^{+} \rightarrow e^{+}\nu(\gamma)$ decay shows a pion stop followed by a MIP track. In the case of the decay channel involving an intermediate muon, there are two particle stops (pion and muon) followed by the MIP. Using cuts on the topography a selection efficiency of about 50\,\% for the direct decay can be reached while rejecting the indirect one by a factor of about 500. Furthermore, the spatial resolution of the ATAR allows to connect the shower in the calorimeter with its corresponding pion stop and by this to resolve pile-up and to reject other positron tracks originating from unrelated muon decays. For the reconstruction of events a traditional rules based approach is followed as well as a more agressive using techiques of artificial intelligence \cite{ref:omar}.

\section{ATAR Sensors}

The sensing element of the ATAR is a trench isolated LGAD strip detector of $2\times 2\,$cm$^2$ size containing 100 micro strips with a pitch of \qty{200}{\micro\metre}. The connection to the readout electronics is provided by long flex cables placed on one end of the strips. Each sensor plane is rotated by $90^\circ$ with respect to its neighbor
to obtain 3D positional track information.
The thickness of the sensors is \qty{120}{\micro\metre}. The inactive Silicon from the support wafer is removed as much as possible (only about \qty{5}{\micro\metre} remain). So the total thickness of the target is about \SI{6.6}{mm}. The sensor thickness has to fulfill several conflicting requirements and is therefore a compromise. It hast to provide sufficient mechanical stability to avoid a support structure and the space between a sensor surface and its fourth successor must be wide enough to mount and connect the flex cable to the readout electronics. Further the signal charge must be high enough to efficiently detect MIPs, also with a moderate gain of ten or below. These conditions favor a higher sensor thickness. At the same time  charge collection has to be fast to satisfy a 2ns pulse pair separation.
The geometry should also give a good spatial resolution along the beam axis and contain several measurement points even for a (low momentum) muon track from the dominant pion decay. The latter conditions favor thin sensors.

Trench isolation was chosen due to its high fill factor and low cross talk to the neighbor channels.

\begin{figure}
    \centering
        \includegraphics[width=0.75\linewidth]{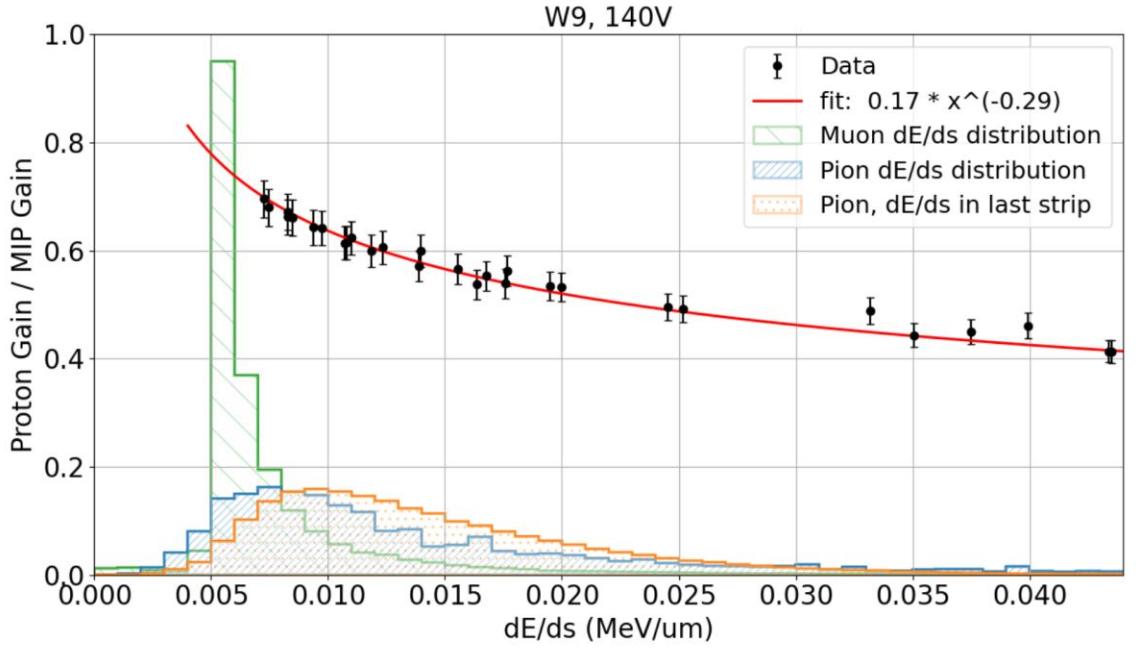}
        \caption{Gain suppression fit \cite{ref:gainSuppressionFig}}
        \label{fig:GainSuppFit}
\end{figure}

As the pulse height of the signals is an important input to the data analysis, the linearity of the signals in respect to the deposited energy has been studied in detail. LGADs show a non-linearity known as gain suppression, which depends on the concentration of the signal charge in the sensor's gain layer and therefore on the total deposited energy and the impact angle of the track. In a series of test beams at the CENPA tandem accelerator this feature has been quantified \cite{ref:gainSuppression}. For tracks traversing the sensor the charge gain follows a power law as indicated in fig.~\ref{fig:GainSuppFit} \cite{ref:gainSuppressionFig}. This relation can be used to correctly calculate the charge deposition. However, this model does not hold true for large, local charge deposits caused by
proton stops as used in the mentioned study. Its behavior with respect to muon stops will be investigated in the end of 2026 testbeam at PSI.

In the present status of prototyping pad and strip LGAD sensors suitable for PIONEER were provided by other projects. The first PIONEER specific production of $2\times 2\,$cm$^2$ (and $1\times 1\,$cm$^2$) strip sensors with \qty{200}{\micro\metre} pitch is ongoing at FBK \cite{ref:FBK} and sensor delivery is expected for early 2027.

\section{ATAR Signal Amplification and Digitization}

To record two subsequent pulses within only 2\,ns in the same channel with a time resolution of the order of 100\,ps, the wave forms of all channels have to be recorded. This requires fast and therefore powerful amplifiers which in consequence need to be cooled. Such a massive infrastructure cannot be placed in front of the calorimeter and is therefore positioned upstream outside the fiducial volume. The connection between strips and preamplifiers is done via $15\,$cm long multi wire flex cables.
%as indicated in fig.~\ref{fig:targetRegion}.
They are attached to the sensors on the one side and to the read out boards on the other.

The amplification of the signals is done using the analog version of the FAST3 chip \cite{ref:fast3} optimized for the fast readout of LGAD sensors. Each chip contains 16 channels (20 in the next version) with adjustable gain. So a read out board serving a full sensor needs to be equipped with five chips. The dynamic range of the amplifier is sufficient to distinguish the signals of MIPs, moving pions and muons and particle stops.

Waveforms containing hits will be digitized. The window for digitization will be open during the time span from about 200\,ns before to about 800\,ns after the pion stop causing the trigger \cite{ref:trigger}.

For the digitization of the waveforms two options are presently followed. One is to sample this whole period with a rate high enough to extract the information (time of pulses, area below the wave forms) with sufficient accuracy and only write the relevant parts of the waveform to tape. This strategy is followed using the HD-SoC chip of Nalu Scientific \cite{ref:HD-SoC}. This chip has 32 channel (64 in the next version), a sampling rate of at 1-2 Gs/s and a buffer length of up to 2048 samples. This allows for sufficiently long wave form to be digitized. Tests with the evaluation board at a sample rate of 1 Gs/s showed a very good linearity. However, two consecutive pulses within 2\,ns might not be resolved with a probability of about \SI{8}{\percent}.

The second option followed is the use of the Sampic system developed by IJCLab Orsay \cite{ref:Sampic}. It has a very high sampling rate of 1.6 to 10\,Gs/s but a short buffer of 64 samples per channel. However, it is possible to daisy chain two or four channels that can re-trigger (even in the same digitization window) without dead time. So up to four subsequent particle pulses per channel within a very short time span can be recorded.

\section{ATAR 8}

\begin{figure}
    \centering
    \begin{subfigure}[c]{0.45\linewidth}
        \centering
        \includegraphics[height=4cm]{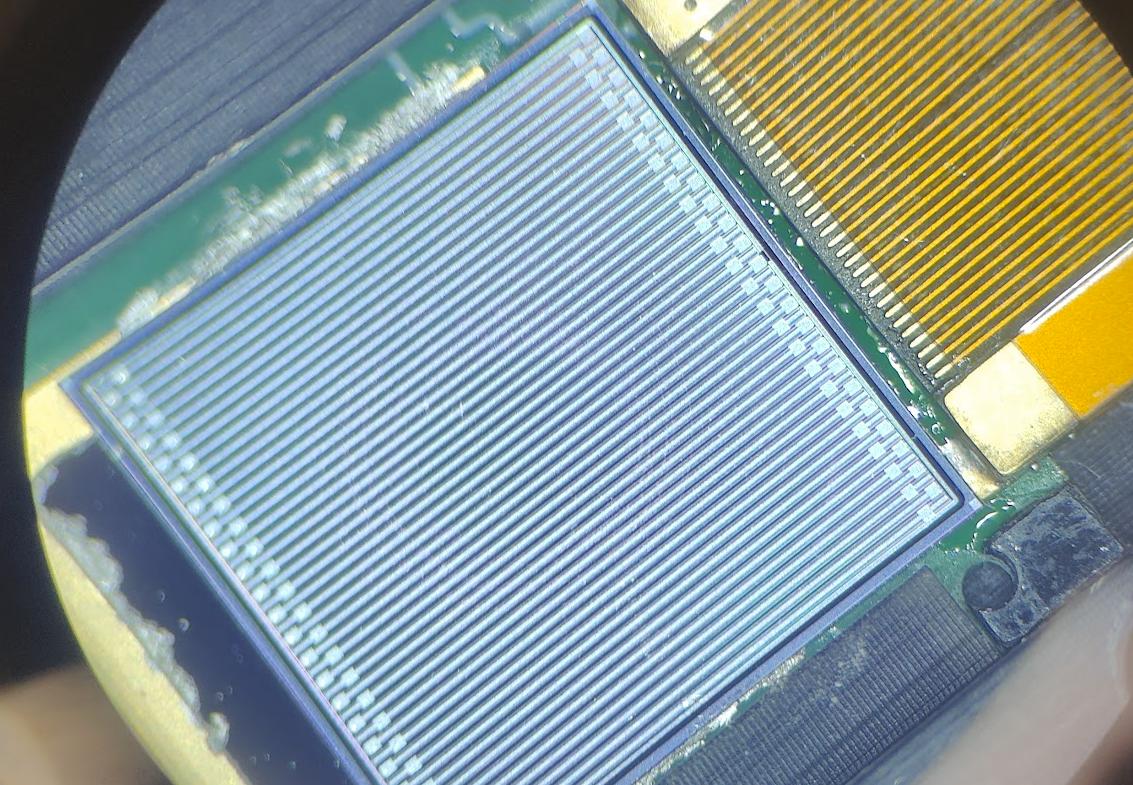}
        \caption{LGAD strip sensor and flex cable mounted on a support frame before wire bonding}
        \label{fig:mountedSensor}
    \end{subfigure}
    \begin{subfigure}[c]{0.45\linewidth}
       \centering
       \includegraphics[height=4cm]{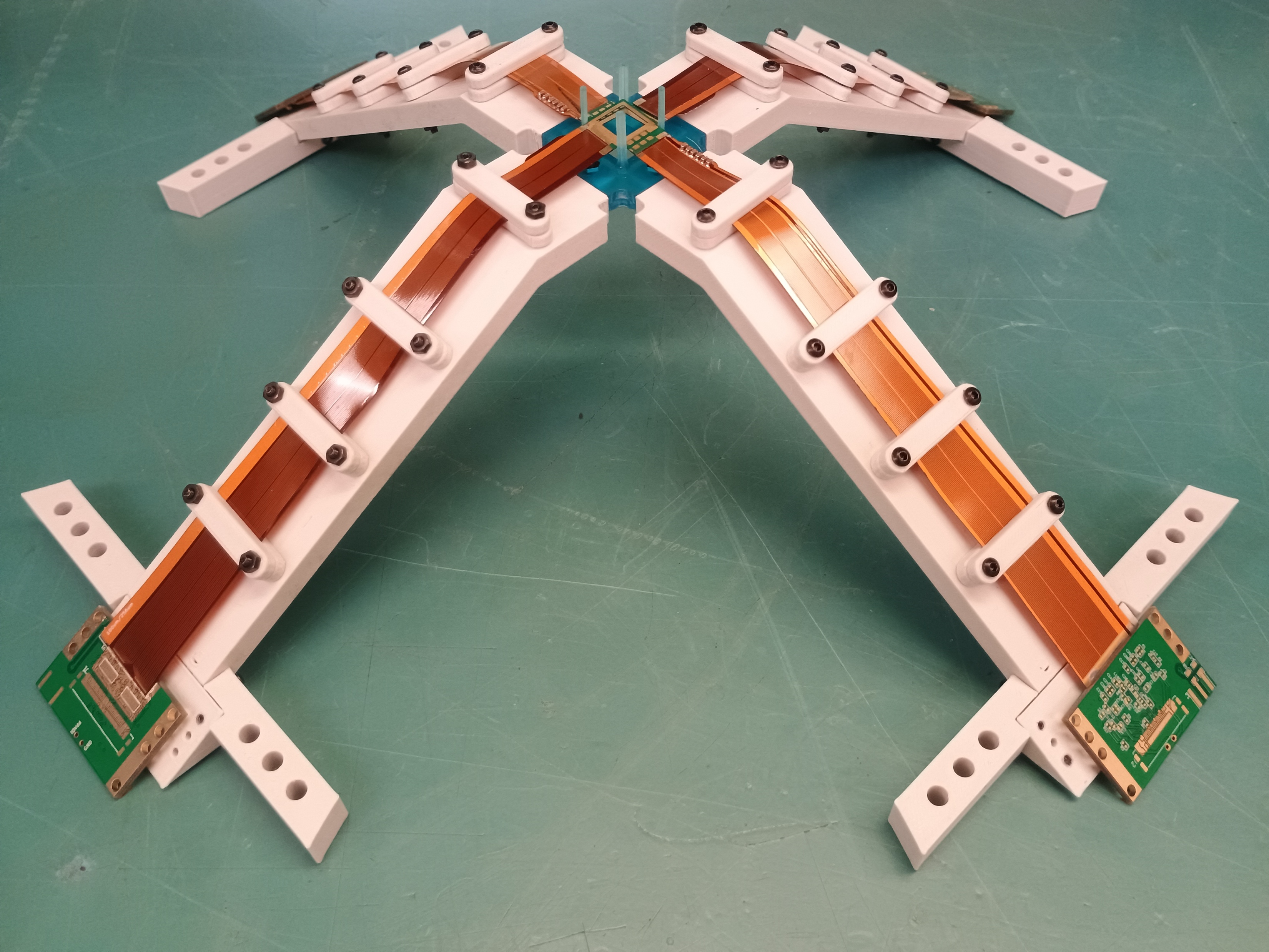}
       \caption{Mechanics for the Atar-8 (mechical design O. Giordiano)}
       \label{fig:8tar_mechanics}
    \end{subfigure}
    \caption{Parts of the ATAR-8 assembly for test beam measurements in Dec.\@ 2026 (credits S.~Mazza, UCSC)}
\end{figure}
To demonstrate the feasibility of ATAR construction and operation a set of prototypes with increasing size will be produced and tested. After a test beam at Triumf where muon decays could be measured in a single detector layer \cite{ref:ultima} the next step is to built ATAR 8, a stack with 8 to 10 sensor layers and to operate it in the PSI-beamline piM1 in Dec.\@ 2026. Sensors used are $10\times 10\,$mm$^2$ in size and contain 48 strips with a pitch of  \qty{200}{\micro\metre}. The readout boards contain two FAST3 chips which 16 channels each. So eight strips on each side of the sensors are not read out but wire bonded to ground. Two of this roughly \qty{120}{\micro\metre} thick sensors are mounted back-to-back in a $90^\circ$-angle onto both faces of a \qty{300}{\micro\metre} thick FR-4 frame which also holds the 15\,cm long flex cables. In order to match the impedance requirements of the FAST3 chip of $100\,\Omega$ these flexes need to have a thickness of \qty{200}{\micro\metre}. The electrical connection between sensor and flex is done via wire bonding. The air gap between the back-to-back layers will be reduced to 5-\qty{10}{\micro\metre} for the final detector. The production of the 4 double layers for ATAR 8 will be finished by end of Oct.\@ 2026.

\section{Conclusion}

The PIONEER experiment aims for a much improved measurement of the  $R^{\pi}_{e/\mu}$ branching ratio. It main components are the highly segmented LYSO crystal calorimeter and the fully active target (ATAR). There will be several ATAR prototypes with increasing size, with the next, ATAR 8, having 8-10 layers to be tested in the beam by the end of 2026. The experiment plans for a staged start-up after the shutdown for the PSI HIPA upgrade.

\section{Acknowledgments}

This work was supported by the United States Department of Energy (DOE), grants DE-SC0010107, and DE-FG02-97ER4102, the Japan Society for the Promotion of Science (JSPS) with KAKENHI grant number 25K24622, and the Swiss National Science Foundation (SNSF), project grant No. 200021\_215301.

% This work was supported by the United States Department of Energy grants DE-FG02-04ER41286336 and DE-FG02-97ER4102 and the Swiss National Science Foundation (SNSF) project grant No.200021\_215301.

\bibliographystyle{JHEP}
\bibliography{rohe_pioneer_vertex-2026_V2}{}

\end{document}